\documentclass[article,balancelastpage,twocolumn,prl]{revtex4}%
\usepackage{makeidx}
\usepackage{amssymb}
\usepackage{dcolumn}
\usepackage{graphicx}
\usepackage{acronym}
\usepackage{amsmath}
\usepackage{amsfonts}%
\ifx\pdfoutput\relax\let\pdfoutput=\undefined\fi
\newcount\msipdfoutput
\ifx\pdfoutput\undefined\else
\ifcase\pdfoutput\else
\ifx\paperwidth\undefined\else
\ifdim\paperheight=0pt\relax\else\pdfpageheight\paperheight\fi
\ifdim\paperwidth=0pt\relax\else\pdfpagewidth\paperwidth\fi

\begin{document}
	
\title{Three-dimensional transformation for rotating coordinate systems and its applications in physics}
\author{B. V. Gisin }

\affiliation{E-mail: borisg2011@bezeqint.net}
\date{\today }
	
\begin{abstract}
Applications of the three-dimensional transformation for rotating coordinate systems to quantum mechanics, general theory relativity and optics are considered.

\textbf{Keywords}: Transformation for point rotating frames, Anomalous magnetic moment, Rotating universe, Single side-band modulator.  
\end{abstract}.
	
\maketitle
	
\section{Introduction}

The recently obtained three-dimensional transformation for rotating coordinate systems can find various applications from micro to macrocosm \cite{pr}. This transformation was obtained by successive rotations of three pairs of differentials of three cylindrical coordinates of the angle, length along the axis of rotation, and time. Such rotations are principally different from  pairs of Cartesian coordinates, since the cylindrical coordinates has a pair that describes the motion along a circle, that is, non-rectilinear movement.

The transformation is designed for describing circularly polarized fields. In such a description the concept of the point-wise rotation is used when the axis of rotation exists at each point. The cylindrical radius in the transformation is an undefined parameter.   

In this paper we describe applications of the transformation to quantum mechanics, general theory of relativity and optics.

\section{The 3D transformation}

It is well known that a three-dimensional transformation in Cartesian coordinates can be obtained by successive rotations of any three pairs of coordinate differentials. If one of the coordinates in the pair is time, then the hyperbolic sine or cosine is used. Such a transformation depends on three arbitrary angles. In contrast, for the cylindrical coordinates, the three angles can be determined uniquely.

One of the angles corresponds to an element of length on a circle of some radius. At a low frequency, rotation has a non-relativistic form \cite{Lan}
\begin{equation}
d\varphi'=d\varphi-\omega dt, \quad dt'=dt. \label{Lan}
\end{equation} 
where $\varphi$ is cylindrical angle, $\omega$ is the angular frequency, $t$ is time. Variables of the second system are primed. 

More modern form of this rotation is
\begin{align}
& rd\varphi^{\prime} =rd\varphi\cosh\Phi-dt\sinh\Phi,\label{Tp1} \\
& dt^{\prime} =-rd\varphi\sinh\Phi+dt\cosh\Phi,\label{Tt1} 
\end{align}
where $\Phi $ is the rotation angle depending on the cylindrical radius $ r $, and we use units with $c=1$. 

Other pairs have the usual rotation, but all angles must also depend on $ r $. In a sense, this is equivalent to the introduction of centrifugal forces.

The denominators of the hyperbolic sine and cosine in (\ref {Tp1}), (\ref{Tt1}) are $\sqrt {1-r^2\omega^2} $. From this follows the well-known assertion that the velocity along a circle of radius $ r $ does not exceed the speed of light. This is equivalent to inequality
\begin{equation} 
r \leq \frac{1}{\omega}=\frac{\lambda}{2\pi}=\lambdabar, \label{rmin}
\end{equation}
where the wavelength $\lambda$ corresponds to the frequency $\omega$.
This important inequality defines the maximum allowable value of $r$.   

Below we briefly consider the derivation of 3D transformation as the result of rotations of the corresponding pairs.

The rotation of first pair (\ref{Tp1}), (\ref{Tt1}) matches a line element $rd\varphi$ on the circle of radius $r$ and time. The angle $\Phi$ is determined by relation
\begin{align}
\tanh\Phi =r\omega. \label{tanh}%
\end{align}

The second is a rotation in the plane of line elements $(dz, rd\varphi')$
\begin{align}
& dz'=dz\cos\Phi_{1}-rd\varphi'\sin\Phi_{1},\label{z1} \\
& rd\tilde{\varphi}=dz\sin\Phi_{1}+rd\varphi'\cos\Phi_{1}, \label{p2}%
\end{align}
orientates the linear element $rd\varphi$ relative the axis of rotation.

The third is Lorentz's transformation of line elements $ dz^{\prime}, dt^{\prime}$
\begin{align}
& d\tilde{z}=dz'\cosh\Phi_{2}-dt'\sinh\Phi_{2}, \label{z2} \\
& d\tilde{t}=-dz'\sinh\Phi_{2}+dt'\cos\Phi_{2}, \label{t2} \\
\end{align}
This transformation is corrected momentum along the new $\tilde{z}$ axis.

3D transformation, as the dependence $rd\tilde{\varphi}, d\tilde{z}, d\tilde{t}$ from $d\varphi, z, t$ can be obtained after excluding the primed coordinates from the relations (\ref{p2}), (\ref{z2}), (\ref{t2}).

The dependence of the angle $\Phi_ {1} $ on $ r $ is determined from the constancy of the speed of light:
\begin{align}
\text{if} \;\; V \equiv \frac{dz}{dt}=1,\;\; \text{then} \;\; \tilde{V} \equiv \frac{d\tilde{z}}{d\tilde{t}}=1, \label{V}
\end{align} 

Using the definition we find
\begin{align}
& \sin\Phi_{1}=\tanh\Phi=r\omega, \quad \cos\Phi=\sqrt{1-r^2\omega^2}, \label{Ph} \\
& \sinh\Phi=\frac{r\omega}{\sqrt{1-r^2\omega^2}}, \quad \cosh\Phi=\frac{1}{\sqrt{1-r^2\omega^2}}. \label{Ph1}
\end{align}

The transformation has a singularity if $ r^2\omega^2 \rightarrow 1 $, since the denominator of sine and cosine is $\sqrt {1-r^2\omega^2} $. However, this singularity can be eliminated by choosing the angle $\Phi_2 $.
\begin{align}
 \exp\Phi_2=\frac{\varkappa}{\sqrt{1-r^2\omega^2}}, \label{Ph2}
\end{align}
where $\varkappa$ is a function of $r$, nonsingular at the point $r\omega=1$. 

Finally, the three dimensional transformation, where the tilde corresponds to the rotating frame, can be written as \cite{Dir}.
\begin{align}
d\tilde{\varphi}  &  =d\varphi+\omega dz-\omega dt,\label{tphi}\\
d\tilde{z}  &  =-\frac{r^{2}\omega}{\varkappa}d\varphi+\kappa_{22}%
dz+\kappa_{23}dt,\label{tz}\\
cd\tilde{t}  &  =-\frac{r^{2}\omega}{\varkappa}d\varphi+\kappa_{32}%
dz+\kappa_{33}dt. \label{tt}%
\end{align}

The inverse transform derived from it algebraically is also used below
\begin{align}
& d\varphi=d\tilde{\varphi}-\frac{\omega}{2\varkappa}d\tilde{z}+ \frac{\omega}{2\varkappa}d\tilde{t},\label{rphi}\\
& dz=r^{2}\frac{\omega}{2} d\tilde{\varphi}+\varkappa_{22}d\tilde{z}-\varkappa_{32}d\tilde{t}, \label{rtz}\\
& dt=r^{2}\frac{\omega}{2}d\tilde{\varphi}-\varkappa_{23}d\tilde{z}+ \varkappa_{33}d\tilde{t}. \label{rtt}%
\end{align}  

The parameter $\kappa_ {kl}$ in components is
\begin{align}
\kappa_{22} & =\frac{1}{2\varkappa}(1+\varkappa^{2}-\omega^{2}r^{2}), \label{C22}\\
\kappa_{23} & =\frac{1}{2\varkappa}(1-\varkappa^{2}+\omega^{2}r^{2}), \label{C23}\\
\kappa_{32} & =\frac{1}{2\varkappa}(1-\varkappa^{2}-\omega^{2}r^{2}), \label{C32}\\
\kappa_{33} & =\frac{1}{2\varkappa}(1+\varkappa^{2}+\omega^{2}r^{2}). \label{C33}%
\end{align}

Along with the speed of light constancy, the transformation possesses following properties:

\textbf{1.} The determinant of system of equation (\ref{tphi}-\ref{tt}) is equal to 1.

\textbf{2.} The transformation turns into the Lorentz transformation if the frequency $\omega$ is equal to zero. 

To do this, it is enough to assume that
\begin{align} 
& \varkappa=\sqrt\frac{1+v}{1-v}, \nonumber  
\end{align}
where $v$ is a normalized velocity. We obtain then
\begin{align}
d\tilde{\varphi}=d\varphi, \quad
d\tilde{z}=\frac{dz-vdt}{\sqrt{1-v^2}}, \quad d\tilde{t}=\frac{-vdz+dt}{\sqrt{1-v^2}}.  \label{LT}
\end{align}

\textbf{3.} There exist two invariants under the 3D transformation. 

This is three-dimensional  quadratic form
\begin{align}
r^2 d\varphi^2+dz^{2}-dt^{2}, \label{ic3}
\end{align}
and quadratic three-dimensional  differential form
\begin{align}
\frac{\partial}{r^2\partial{\varphi^2}}+\frac{\partial}{\partial{z^2}}- \frac{\partial}{\partial{t^2}}, \label{id3}
\end{align}

The invariance of the form (\ref {ic3}) leads to the important conclusion that the interval of the flat Minkowski space is also invariant with respect to 3D transformation. This does not exclude possibility that space can be an object of point rotation, that is, 'matter of free space' rotates similarly to an electromagnetic field in a circularly polarized wave.

Consider applications of 3D transformation in quantum mechanics.

\subsection{Application to quantum mechanics}

 Consider Dirac's equation in constant and rotating magnetic field, the plane of rotation is perpendicular to direction of the constant field. 
 
 This equation is not invariant under the 3D transformation. Consequently, the equation transformed into a rotating reference system along with the wave function is supplemented by a new term depending on angles $\Phi, \Phi_1, \Phi_2$ and their derivatives in respect to $r$. 
 
 The angle $\Phi_2$ is defined with help of a differential equation of the first order \cite{Dir}. This makes it possible removal singularities at the point $r^2\omega=1$ 
\begin{equation}
\Phi_{2,r}=\Phi_{,r}\sin\Phi_{1},\text{ }\exp\Phi_{2}=\frac{\varkappa}%
{\sqrt{1-\Omega^{2}r^{2}/c^{2}}}. \label{f2}%
\end{equation}
The dependence of $\Phi_2$ has the same form as in (\ref{Ph2}), however, unlike to this definition, $\varkappa$ as the constant of integration is not depend on $r$. 

This dependence significantly simplifies the new term, which take the form
\begin{align}
& -(1-\alpha_{3})i\alpha_{1}\alpha_{2}\frac{\hbar\omega}
{2\varkappa}\tilde{\Psi}, \label{add}
\end{align}
where $\omega$ is the frequency of the oscillating magnetic field,  $\tilde{\Psi}$ is the transformed wave function.

Description of the magnetic resonance gives an argument in favor of the 3D transformation. The resonance occurs when a particle moves in a strong constant and a weak oscillating magnetic field. From the point of view of symmetry, a rotating field is more acceptable, but creating such a field is more difficult. However, the oscillatory field can be represented as the sum of two opposite rotating fields. The effect of one of them is negligible and can be excluded from consideration.

The classical description of spin components in a rotating magnetic field implies a transition to a static problem, that is, to a coordinate system where the magnetic field is at rest. The first step in this description is the non-relativistic transformation of the wave function into a rotating coordinate system $\Psi \rightarrow \exp(i\varphi t/2)\Psi_0 $ \cite {Lan}.

In \cite{Dir} it was shown that the use of a three-dimensional transformation instead of this non-relativistic one and the determination of magnetic resonance as stationarity solutions in a rotating frame of reference leads to the conclusion that the magnetic moment is always anomalous and its $g$-factor equals $\varkappa$.

Such a description cannot give a concrete value of the $g$-factor, this value is calculated by means of quantum field theory with great accuracy.

The study of the additional term (\ref{add}) gives another argument in favor of the three-dimensional transformation. This is related to the invariant form found above (\ref {ic3}) and the possible interpretation of the space as an object of the point rotation.

Suppose that our space is such an object, rotating at a frequency of $\omega_r $. Then, after the transition from the resting space, the Dirac equation gets the term (\ref{add}), where $\omega=\omega_r $, for definiteness $\omega_r$ is named the relic frequency.

In \cite{Dir} an upper boundary for this frequency is found as uncertainty of
Lamb's shift in the hydrogen atom. This boundary is of the order of 14 KHz. In reality this frequency is considerably lesser and correspond cosmic phenomena. 

\subsection{Application to the general theory relativity}

Implementing this idea, insert the invariant (\ref{ic3}) into the interval in the general theory of relativity
\begin{align}
ds^2=fdr^2+g(r^2 d\varphi^2+dz^{2}-dt^{2}), \label{ig}
\end{align}
where $f$ and $g$ are some functions of coordinates $\varphi, z, t$.

Amazingly, calculations show that Einstein's equations with the cosmological term (consisting of 10 equations) have no a solution depending only on the cylindrical radius. However, there is an exact solution for the metric (\ref {ig}) \cite {pr}, compatible with the two functions $ f $ and $ g $. Its distinguishing feature is that the solutions do not depend on $ r $ and represent point rotations with the axis of rotation at each point.

The solution has the form
\begin{align} 
f=\frac{1}{(\nu t+kz)^2}, \quad g=\frac{(\eta\varphi)^2}{(\nu t+kz)^2},\label{fnu}
\end{align}
where $\nu,k$ are constants that satisfy
\begin{align} 
\nu^2-k^2=\frac{1}{3}\Lambda, \label{nk}
\end{align}
$\Lambda$ is the cosmological constant.

The solution has singular point, which moves along the axis of rotation in course of time.

This solution is an example of the 'point rotation'. Together with this solution, a dualism arises, similar the 'wave-particle' in quantum theory. 
On one hand, the size of the universe is bounded according to the inequality (\ref{rmin}), where $1/\lambdabar$ is proportional to the square root of the cosmological constant, on the other hand, the size is infinite, since the axis of rotation can be chosen at each point.

\subsection{Application to optics}

Consider the propagation of an optical circularly polarized wave through a single-band modulator. This is an interesting object for researches, since the movement of light through a modulator is in some sense similar to the movement of spin in magnetic resonance.

The modulator is designed for optical frequency shifting without harmonics. It consists of a trigonal electro-optical crystal in an electric field, rotating in a plane perpendicular to axis of crystal \cite{pat}. 

The principle of operation of the modulator tested experimentally in \cite{Cam}.

Three-dimensional transformation allows us to consider the composition of frequencies as a composition of velocities in the Lorentz transformation.

'On fingers', the process of light movement is described as follows. Let $\omega $ and $\Omega $ be the frequency of the applied electric field and a circularly polarized optical wave, respectively. 
Assume for definiteness that sense of both rotations coincides. We also assume for generality that crystal may move along the crystal axis at a velocity $v$.

Transit into a frame connected with the rotating indicatrix. Then the optical frequency becomes $\Omega-\omega/2$. If the amplitude of the applied electric field is equal to the half-wave value, then the sense of the rotation of the optical frequency reverses and at the output of the crystal in the initial (laboratory) frame takes the value $ -\Omega+\omega $.

More general results can be obtained with help of the frequencies composition in 3D transformation. 

Transit to the rotating frame using relations (\ref{tphi})-(\ref{tt}) and take into account that $\omega$ must be replaced by $\omega/2$ 
\begin{align}
& \tilde{\Omega}=\frac{\varkappa}{\tau}[2\Omega+\omega v-\omega], \nonumber \\
& \tilde{v}=\frac{-r^{2}\omega\Omega+c_{22}v+c_{23}}{-r^{2}\omega\Omega+ c_{32}v+c_{33}}, \nonumber \\
& \frac{d\tilde{t}}{dt}=\frac{\tau}{2\varkappa}, \;\; \tau=[-r^{2}\omega\Omega+ c_{32}v+c_{33}]. \nonumber
\end{align}
$c_{kl}=2\varkappa\kappa_{kl}$ is a matrix inserted for convenience.

Subtract 1 from every site of the second equation.
\begin{align}
& \tilde{\Omega}=\frac{\varkappa}{\tau}[2\Omega+\omega(v-1)], \label{tQQ} \\
& \tilde{v}-1=\frac{2\varkappa^2(v-1)}{\tau}, \quad \tau=\frac{2\varkappa^2(v-1)}{\tilde{v}-1} \label{tVV} \\ 
& \frac{d\tilde{t}}{dt}=\frac{\tau}{2\varkappa}, \;\; \tau=-r^{2}\omega\Omega+ c_{32}(v-1)+2.  \label{tau}
\end{align}
 
The reverse motion is realized with help the inverse transformation.
\begin{align}
& \Omega=\frac{1}{\tilde{\tau}}[2\varkappa\tilde{\Omega}-\omega(\tilde{v}-1)], \label{QtQ} \\
& v-1=\frac{2(\tilde{v}-1)}{\tilde\tau}, \quad \tilde\tau=\frac{2(\tilde{v}-1)}{v-1} \label{VtV} \\
& \frac{dt}{d\tilde{t}}=\frac{\tilde{\tau}}{2\varkappa}, \quad \tilde{\tau}=r^{2}\omega\varkappa\tilde{\Omega}-c_{23}(\tilde{v}-1)+ 2\varkappa^2 \label{ttau} \\
& \tilde{\tau}\tau=4\varkappa^2, \nonumber
\end{align}

Obviously, the system of equations derived from the inverse transformation is a consequence of the direct transformation, in accordance with the fact that both systems follows one from another.

Input parameters at the entrance of the optical wave to the modulator are $\Omega, v$ is transformed to $\tilde{\Omega}, \tilde{v}$. Without modulating by the electric field we obtain the initial parameters at output using the inverse transformation. However, if an amplitude of the applied electric field is equal to the half-wave voltage, then we should insert $-\tilde{\Omega}$ as input parameter into the inverse transformation.  

This phenomenological approach gives the final output frequency. 
\begin{align}
-\Omega+\omega(1-v).\label{Om}
\end{align}
It is noteworthy that the parameter $\varkappa $ vanishes from the final result.  

\section{Conclusion}
The arguments in favor of the tree-dimensional transformation for rotating frames are:

Presently quantum mechanics gives such an argument. This is the statement that in Dirac's equation the magnetic moment is always anomalous. However, it is necessary to redefine the magnetic resonance as the condition of stationarity solution in a rotating frame. In this case the $g$-factor is equal to $\varkappa$

Another more strong argument gives the general theory of relativity. The use of the invariant quadratic form in the interval lead to amazing results. There are no solutions to the Einstein equation with the cosmological term depending only on the cylindrical radius, but there exists the solution with $f$ and $g$ depending on all other coordinates. 

Three-dimensional transformation for a rotating coordinate system allows us to consider the composition of frequencies similar to the composition of velocities in the Lorentz transformation. An argument in favor of 3D transformation is simple summation if the crystal is at rest. Perhaps for a moving crystal, the shift of frequency can be used in the technique.

\end{document}